\documentstyle[12pt]{article}
\begin{document}

\def\three_j(#1,#2,#3,#4,#5,#6){\pmatrix{#1 & #2 & #3\cr
					 #4 & #5 & #6\cr}}

\def\qqq{\end{document}}
\def\pmb#1{\setbox0=\hbox{$#1$}%
\kern-.025em\copy0\kern-\wd0
\kern.05em\copy0\kern-\wd0
\kern-.025em\raise.0433em\box0 }
\def\b{\pmb}

\def\xara(#1,#2,#3,#4){\left(\matrix{#1 & #2\cr #3 & #4\cr}\right)}
\def\Fn{J_{(n)}}
\def\Fs{J_{(s)}}
\def\Hn{H_{(n)}}
\def\Hs{H_{(s)}}
\def\thru#1{\mathrel{\mathop{#1\!\!\!/}}}
\def\CN{\cal N}
\def\w{\omega}
\def\W{\Omega}
\def\six_j(#1,#2,#3,#4,#5,#6){\left\{\matrix{#1 & #2 & #3\cr
					 #4 & #5 & #6\cr}\right\}}
\def\nine_j(#1,#2,#3,#4,#5,#6,#7,#8,#9){\left\{\matrix{#1 & #2 & #3\cr
					#4 & #5 & #6\cr
					 #7 & #8 & #9\cr}\right\}}
\def\W{\Omega}
\def\pd#1#2{{\partial #1\over \partial #2}}
\def\v#1{ {\bf #1} }
\def\Ener(#1,#2){ \sqrt{{#1}^2+{#2}^2} }
\def\c#1{ {\cal #1}}%

\def\overlay#1#2{\setbox0=\hbox{$#1$}\setbox1=\hbox to \wd0{\hss$#2$\hss}#1%
\hskip -1\wd0\copy1}
\newcommand{\xslash}[1]{\overlay{#1}{/}}
\newcommand{\undsim}[1]{\olay{#1}{\sim}}

\def\bold#1{\setbox0=\hbox{$#1$}%
      \kern-.025em\copy0\kern-\wd0
      \kern.05em\copy0\kern-\wd0
      \kern-.025em\raise.0433em\box0 }
\def\tr{\, \hbox{tr} \, }
\def\Tr{\, \hbox{Tr} \, }
\def\sgn{\, \hbox{sgn} }
\def\bra{\langle}
\def\ket{\rangle}
\def\shalf{\,1/2\,}
\def\half{\, {1 \over 2} \,}
\def\gsim{\displaystyle\mathop{>}_{\sim}}
\def\lsim{\displaystyle\mathop{<}_{\sim}}
\def\rd{\partial}
\def\c{\hbox{c}}
\def\s{\hbox{s}}

\def\S11{S_{11}(1535)}
\def\E0+{E_{0^+}}
\def\etaNN{\eta NN^*}
\def\geta{g_{\eta NN^*}}

\def\tdotr{\, \vec \tau \cdot \hat r \, }
\def\fpi{f_\pi}

%%%%%%%%%%%%%%%%%%%%
\renewcommand{\thefootnote}{\fnsymbol{footnote}}
\def\footnoterule{\kern-3pt \hrule width \hsize \kern2.6pt}

%%%%%%%%%%%%%%%%%%%%%%%%%%%%%%%%%%%%%%%%%%%%%%%%%%%%%%%%%%%%%%
%%%%%%%%%%%%%%%%%%%%%%%%%%%%%%%%%%%%%%%%%%%%%%%%%%%%%%%%%%%%%%
%%%%%%%%%%%%%%%%%%%%%%%%%%%%%%%%%%%%%%%%%%%%%%%%%%%%%%%%%%%%%%

\newcommand{\beq}{\begin{equation}}
\newcommand{\eeq}{\end{equation}}
\newcommand{\ba}{\begin{array}}
\newcommand{\ea}{\end{array}}
\newcommand{\beqa}{\begin{eqnarray}}
\newcommand{\eeqa}{\end{eqnarray}}
\newcommand{\bd}[1]{ \mbox{\boldmath $#1$}  }
\newcommand{\sla}[1]{  #1 \!\!\!\slash  }
\newcommand{\tresj}[6]{\left(\!\!\!
			    \begin{array}{ccc}
				      #1&#2&#3\\
				      #4&#5&#6
			    \end{array}\!\!\!
		       \right)}
\newcommand{\seisj}[6]{\left\{\!\!\!
			    \begin{array}{ccc}
				      #1&#2&#3\\
				      #4&#5&#6
			    \end{array}\!\!\!
		       \right\}}
\newcommand{\nuevej}[9]{\left\{\!\!\!
			    \begin{array}{ccc}
				      #1&#2&#3\\
				      #4&#5&#6\\
				      #7&#8&#9
			    \end{array}\!\!\!
		       \right\}}

%%%%%%%%%%%%%%%%%%%%%%%%%%%%%%%%%%%%%%%%%%%%%%%%%%%%%%%%%%%%%%%%%
%%%%%%%%%%%%%%%%%%%%%%%%%%%%%%%%%%%%%%%%%%%%%%%%%%%%%%%%%%%%%%%%
%%%%%%%%%%%%%%%%%%%%%%%%%%%%%%%%%%%%%%%%%%%%%%%%%%%%%%%%%%%%%%%%%%

\setcounter{footnote}{0}
\begin{center}
{\bf SPIN DEPENDENT MOMENTUM DISTRIBUTIONS IN DEFORMED NUCLEI} \\

\vspace*{1cm}
J. A. Caballero, E. Garrido, E. Moya de Guerra, P. Sarriguren,
and J. M. Ud\'{\i}as \\

\vspace*{0.8cm}

	 Instituto de Estructura de la Materia, CSIC \\
	 Serrano 123, 28006 Madrid, SPAIN \\

\vspace*{0.5cm}
\end{center}

\vfill

\begin{abstract}

We study the properties of the spin dependent one body density in
momentum space for odd--A polarized deformed nuclei within the mean field
approximation. We derive analytic expressions connecting intrinsic and
laboratory momentum distributions. The latter are related to
observable transition densities in {\bf p}--space that can be probed in
one nucleon knock--out reactions from polarized targets. It is shown that
most of the information contained in the intrinsic spin dependent
momentum distribution is lost when the nucleus is not polarized.
Results are presented and discussed for two prolate
nuclei, $^{21}$Ne and $^{25}$Mg, and for one oblate nucleus, $^{37}$Ar.
The effects of deformation are highlighted by comparison to the case of
odd--A nuclei in the spherical model.

\end{abstract}
\vspace{0.8cm}
\begin{center}
(25 manuscript pages, 2 tables, and 6 figures)
\end{center}

\vfill

\eject

\vspace{2.0cm}
\noindent
Proposed running head:

\begin{center}
SPIN DEPENDENT MOMENTUM DISTRIBUTIONS
\end{center}
\vspace{4.0cm}
Please send all the correspondence to:

\begin{center}
Prof. E. Moya de Guerra \\
Instituto de Estructura de la Materia \\
Consejo Superior de Investigaciones Cient\'{\i}ficas (CSIC) \\
Serrano 123, MADRID--28006, SPAIN \\
Tel. 34 - 1 - 585 5395 \,\,\,\,\,\,\,\, Fax. 34 - 1 - 585 5184 \\
e--mail: EMELVIRA@IEM.CSIC.ES
\end{center}

\vfill
\eject
%%%%%%%%%%%%%%%%%%%%%%%%%%%%%%%%%%%%%%%%%%%%

%%%%%%%%%%%%%%%%%%%%%%%%%%%%%%%%%%%%%%%%%%%%

\section*{I. Introduction}

Momentum distributions in nuclei are a subject of increasing interest at
both nucleon and parton levels. Reviews of theory and models on nucleon
momentum distributions can be found in Refs.~\cite{AHP88,MaS91}. Experimental
information on nucleon momentum distributions can be gained
from one nucleon knock--out in nuclear photo--absortion, as well as by
$\hbox{A}(\hbox{x},\hbox{x}'\hbox{N})\hbox{B}$
reactions$^{\mbox{\scriptsize\cite{Fes92}}}$ --where a
hadronic or leptonic beam, $\hbox{x}$, scatters and knocks out one nucleon from
a
nuclear target, $\hbox{A}$. As pointed out in Ref.~\cite{AHP88} the
interpretation of information on momentum distributions is not free from
ambiguities because, at present, neither experimental data nor theoretical
considerations are sufficiently comprehensive. Yet fairly reliable
information is already available on nucleon momentum distributions from
electron scattering in the quasi--elastic
region$^{\mbox{\scriptsize\cite{FM84}}}$, as well
as on parton distributions from lepton scattering in the deep inelastic
region$^{\mbox{\scriptsize\cite{Mu90}}}$. Most of the work on complex nuclei
has been
devoted to spherical nuclei, but in the last years several questions have
emerged concerning the role of nuclear deformation in quasi--elastic
electron scattering and in deep inelastic scattering. In the latter, it has
been suggested to use polarized deformed nuclei to study the degree of
transparency in different directions$^{\mbox{\scriptsize\cite{Ja88}}}$;
in the first it has been questioned whether coincidence measurements may
be sensitive to nuclear deformation.

Investigations of quasi--elastic electron scattering from even--even
deformed nuclei have been carried out both theoretically and
experimentally. Starting from mean field calculations with increasing
degree of sophistication, we studied the properties of momentum
distributions in the intrinsic ground state of several even--even axially
symmetric deformed nuclei. We investigated not only effects of deformation
but also effects of pairing and of short range
correlations$^{\mbox{\scriptsize\cite{MCS88,CM90,MSC91}}}$. In a search for
deformation
effects, coincidence (e,e$'$p) experiments on spherical and deformed
Nd isotopes were performed at NIKHEF$^{\mbox{\scriptsize\cite{La93}}}$.
More recently we
have also initiated studies of deformation effects in quasi--elastic
electron scattering from polarized deformed
nuclei$^{\mbox{\scriptsize\cite{CDP94}}}$.
With progress made in developing polarized beams and targets, as well as
in polarimeters designs (in addition to that on CW linac facilities), the
study of spin degrees of freedom is becoming a major theme in Nuclear
Physics either at low, intermediate or high energies. In this paper we
attempt to lay a basis for such studies in odd--A deformed nuclei.

In general the one--body density (either in {\bf r}--space or in {\bf
p}--space) is a two by two matrix in spin space. For closed shell (spin
saturated) nuclei the one--body density is proportional to the unit matrix
in spin space, i.e., the momentum distribution is independent of spin. The
same is true for even--even axially symmetric deformed nuclei in the mean
field approximation due to time reversal invariance of the intrinsic
ground state. However, for an odd--A deformed nucleus the intrinsic ground
state is no longer time reversal invariant and the one--body density
matrix is in general non diagonal in spin space. Hence {\bf scalar} and
{\bf vector} momentum distributions can be defined in the intrinsic frame,
that contain all the information on the intrinsic spin dependent momentum
distribution. This paper is devoted to the study of these new momentum
distributions.

It is important to realize that in order to access to this new information
one needs to relate the observable spin dependent momentum distributions
of the polarized nucleus in the laboratory frame with the intrinsic scalar
and vector momentum distributions. A central issue of this paper is
to establish such relations.

We study properties of scalar and vector momentum distributions within the
self--consistent mean field approximations assuming axially symmetric and
time--reversal invariant mean fields. To this end we present and discuss
results of density dependent Hartree--Fock (DDHF) calculations for two
nuclei $^{21}$Ne and $^{37}$Ar that have ground states with equal spin and
parity ($3/2^+$) but have quite different intrinsic structure. Results for
$^{25}$Mg
are also discussed. The self--consistent mean fields for $^{21}$Ne and
$^{25}$Mg are prolate whereas that for $^{37}$Ar is oblate, and the
$K$--values are different for each nucleus. The role of deformation is
further highlighted by comparison to the limit of a spherical mean field,
i.e., to the case of odd--A nuclei with a single--nucleon outside closed
shells.

The paper is organized as follows: in Section II the intrinsic scalar and
vector momentum distributions are defined and their properties are
studied. In Section III we define scalar and vector momentum distributions
for the polarized nucleus in the laboratory frame, we establish their
connection with the intrinsic ones, as well as with observable transition
densities in {\bf p}--space. In this section the spherical limit is also
discussed and comparisons with results on the deformed nuclei are
presented. Section IV summarizes our conclusions.

%%%%%%%%%%%%%%%%%%%%%%%%%%%%%%%%%%%%%%%%

\section*{II. Intrinsic Momentum Distributions}

To study spin dependent momentum distributions in axially symmetric deformed
nuclei our starting point are the single--particle Hartree--Fock (HF) solutions
in
coordinate space, which are characterized by the angular momentum projection
along the symmetry axis ($\Omega_i$) and by the parity ($\pi_i$). The HF
wave functions are expressed in the intrinsic system with the $z$ axis
the symmetry axis and the ($x,y$) plane the symmetry plane. The states $i$,
$\bar{i}$ degenerate in energy are written
as$^{\mbox{\scriptsize\cite{VB72}}}$

\beq
\Phi_i(\bd{r},s,q)=\chi_{q_i}(q)\left[\phi_i^{(+)}(r_\perp ,z)
		e^{i\varphi \Lambda^-} \chi_{+}(s) +
		\phi_i^{(-)}(r_\perp ,z)
		e^{i\varphi \Lambda^+} \chi_{-}(s)\right] ,
		\label{1}
\eeq
\beq
\Phi_{\bar{i}}(\bd{r},s,q)=\chi_{q_i}(q)\left[-\phi_i^{(-)}(r_\perp ,z)
		e^{-i\varphi \Lambda^+} \chi_{+}(s) +
		\phi_i^{(+)}(r_\perp ,z)
		e^{-i\varphi \Lambda^-} \chi_{-}(s)\right]
		\label{2}
\eeq
where $\chi_{q_i}(q)$, $\chi_{\pm}(s)$ are isospin and spin functions,
$\bar{i}$ is the time reverse of $i$, $\Lambda^{\pm}=\Omega_i \pm 1/2
\ge 0$, and $r_{\perp}$, $z$, $\varphi$ are the cylindrical coordinates of
$\bd{r}$. The wave functions $\phi_i^{(\pm)}$ are expanded into eigenfunctions
of an axially deformed Harmonic Oscillator, characterized by the set of
quantum numbers $\{\alpha\}=\{n_{\perp},n_z,\Lambda ,\Sigma \}$, with expansion
coefficients $C_{\alpha}^i$. Taking the Fourier transforms of Eqs.~(1) and (2)
we get the single--particle HF wave functions in momentum space

\beq
\tilde{\Phi}_i(\bd{p},s)=\tilde{\phi}_i^{(+)}(p_\perp ,p_z)
		e^{i\varphi_p \Lambda^-} \chi_{+}(s) +
		\tilde{\phi}_i^{(-)}(p_\perp ,p_z)
		e^{i\varphi_p \Lambda^+} \chi_{-}(s) ,
		\label{3}
\eeq

\beq
\tilde{\Phi}_{\bar{i}}(\bd{p},s)=\left[-\tilde{\phi}_i^{(-)}(p_\perp ,p_z)
		e^{-i\varphi_p \Lambda^+} \chi_{+}(s) +
		\tilde{\phi}_i^{(+)}(p_\perp ,p_z)
		e^{-i\varphi_p \Lambda^-} \chi_{-}(s)\right]\pi_i
		\label{4}
\eeq
with $p_{\perp}$, $p_z$, $\varphi_p$ the cylindrical coordinates of $\bd{p}$.
Here and in what follows we omit isospin to abbreviate the notation. The
wave functions $\tilde{\phi}_i^{(\pm)}$ are given by

\beq
\tilde{\phi}_i^{(\mp)}(p_{\perp},p_z)=\sum_{n_\perp ,n_z}C^{i}_{\alpha^{\pm}}
		\tilde{\phi}_{\alpha^{\pm}}(p_{\perp},p_z)
	\label{5}
\eeq
where $\alpha^{\pm}$ stands for $\{n_{\perp}$, $n_z$, $\Lambda=
\Lambda^{\pm}\}$ and
the sum extends over basis states with major quantum numbers
($N=2n_{\perp}+n_z+\Lambda$) even or odd for $\pi_i=+1$ and $\pi_i=-1$,
respectively. In the cylindrical basis

\beq
\tilde{\phi}_{\alpha^{\pm}}(p_{\perp},p_z)=\frac{(-i)^N\sqrt{\pi_i}}
				{\sqrt{2\pi}}\psi_{n_\perp}^{\Lambda^\pm}
			(p_{\perp})\psi_{n_z}(p_z)
	\label{6}
\eeq
with

\beq
\psi_{n_\perp}^{\Lambda}(p_{\perp})=
		\left[
			\frac{2n_\perp !}{\beta^2_\perp (n_\perp +\Lambda)!}
		\right]^{1/2}
			\eta^{\Lambda/2}e^{-\eta/2}L^{\Lambda}_{n_\perp}(\eta) ,
	\label{7}
\eeq

\beq
\psi_{n_z}(p_z)=
		\left[
			\frac{1}{\beta_z \sqrt{\pi} 2^{n_z}n_z !}
		\right]^{1/2}
			e^{-\xi^2 /2}H_{n_z}(\xi)
	\label{8}
\eeq
with $L_{n_\perp}^{\Lambda}$ and $H_{n_z}$ associated Laguerre and Hermite
polynomials respectively, $\eta=p_{\perp}^2/\beta_{\perp}^2$,
$\xi=p_z/\beta_z$ and $\beta_\perp$, $\beta_z$ the inverse of the harmonic
oscillator length
parameters ($\beta_\perp = (m\omega_\perp /\hbar)^{1/2}$,
$\beta_z=(m\omega_z/\hbar)^{1/2}$).
For later use it is convenient to expand the momentum
dependent single--particle wave functions in a spherical basis. We write

\beq
\tilde{\Phi}_i(\bd{p},s)=\sum_{\ell j} {\cal Y}_{\ell j}^{\Omega_i}(
		\hat{p},s)\tilde{\phi}_i^{\ell j}(p)
	\label{9}
\eeq

\beq
\tilde{\Phi}_{\bar{i}}(\bd{p},s)=
		\sum_{\ell j}(-1)^{\Omega_i-j}
		{\cal Y}_{\ell j}^{-\Omega_i}(
		\hat{p},s)\tilde{\phi}_i^{\ell j}(p)
	\label{10}
\eeq
with

\beq
{\cal Y}_{\ell j}^m(\hat{p},s)=\sum_{\Lambda \Sigma}
		\langle \ell\,\, \Lambda\,\, \frac{1}{2}\,\, \Sigma |
		j\,\, m \rangle
		Y^{\Lambda}_{\ell}(\hat{p})\chi_{\Sigma}(s)
	\label{11}
\eeq
and

\beq
\tilde{\phi}^{\ell j}_{i}(p)=\sum_{\alpha}C_{\alpha}^i
	\langle \ell\,\, \Lambda \,\,\frac{1}{2}\,\, \Sigma |j\,\, \Omega_i \rangle
		R^{\alpha}_{\ell}(p)
	\label{12}
\eeq

\beqa
& & R^{\alpha}_{\ell}(p)=\int d\Omega_p Y^{\Lambda \ast}_{\ell}(\Omega_p)
	\tilde{\phi}_{\alpha}(p_\perp ,p_z)e^{i\varphi_p \Lambda}
	\nonumber \\
&\!\!\!\!=& \!\!\!(-i)^{N+2\Lambda}\sqrt{\pi_i}
	\left[\frac{(2\ell+1)(\ell -\Lambda)!}{2(\ell +\Lambda)!}\right]
		^{1/2}
	\int_0^{\pi}\sin\theta_p d\theta_p P_\ell ^\Lambda (\cos \theta_p)
	\psi_{n_\perp}^\Lambda(p \sin\theta_p)\psi_{n_z}(p\cos \theta_p)
	\nonumber \, .\\
 &  &   \label{13}
\eeqa

We represent by $\Psi_k$ the intrinsic ground state of an odd--A nucleus
with odd nucleon in state $k$. $\tilde{\Phi}_k$ represents the wave function in
momentum space for the state $k$, and $\pi_k$, $\Omega_k=K$ are its quantum
numbers.
The intrinsic one--body spin dependent momentum
distribution is then given by

\beqa
& & M_{\Sigma \Sigma'}(\bd{p})=\langle\Psi_k|
	a^+_{\bd{p}\Sigma}a_{\bd{p}\Sigma'}|\Psi_k\rangle
 \nonumber \\
& =& \sum_{i\neq k}v_i^2\left[
	\langle \chi_{\Sigma'}|\tilde{\Phi}_i(\bd{p},s)\rangle
	\langle \tilde{\Phi}_i(\bd{p},s)|\chi_{\Sigma}\rangle +
	\langle \chi_{\Sigma'}|\tilde{\Phi}_{\bar{i}}(\bd{p},s)\rangle
	\langle \tilde{\Phi}_{\bar{i}}(\bd{p},s)|\chi_{\Sigma}\rangle
	\right]
\nonumber \\
& + &\langle \chi_{\Sigma'}|\tilde{\Phi}_k(\bd{p},s)\rangle
	\langle\tilde{\Phi}_k(\bd{p},s)|\chi_{\Sigma}\rangle
\label{14}
\eeqa
with $v_i^2$ occupation probabilities. The last term in this equation gives
the contribution from the odd nucleon, while the first term gives the
contribution from the nucleons in the even--even core.
Decomposing this matrix into a {\bf scalar} ($\overline{M}$) and a {\bf vector}
($\widehat{\bd{M}}$) in spin space

\beqa
\overline{M} & = & Tr\left[M(\bd{p})\right] \\
\widehat{\bd{M}} &=& Tr\left[M(\bd{p})\bd{\sigma}\right]
\label{16}
\eeqa
it is easy to show that the even--even core contributes only to the scalar
momentum distribution,

\beqa
\overline{M} (\bd{p})& = & \sum_{i\neq k} 2v_i^2\left(
	|\tilde{\phi}_i^{(+)}(p_\perp ,p_z)|^2 +
	|\tilde{\phi}_i^{(-)}(p_\perp ,p_z)|^2
		\right)
\nonumber \\
& +&    |\tilde{\phi}_k^{(+)}(p_\perp ,p_z)|^2 +
	|\tilde{\phi}_k^{(-)}(p_\perp ,p_z)|^2 ,
	\label{17}
\eeqa
while the vector momentum distribution is made up
entirely by the odd nucleon,

\beq
\widehat{M}_z(\bd{p})=
	|\tilde{\phi}_k^{(+)}(p_\perp ,p_z)|^2 -
	|\tilde{\phi}_k^{(-)}(p_\perp ,p_z)|^2
\label{18}
\eeq

\beq
\widehat{M}_x(\bd{p})\equiv \cos\varphi_p \widehat{M}_\perp (p_\perp ,p_z)
		\label{19}
\eeq

\beq
\widehat{M}_y(\bd{p})\equiv \sin\varphi_p \widehat{M}_\perp (p_\perp ,p_z)
		\label{20}
\eeq
with

\beq
\widehat{M}_\perp (p_\perp ,p_z)=2\tilde{\phi}_k^{(+)}(p_\perp ,p_z)
	\tilde{\phi}_k^{(-)}(p_\perp ,p_z)  \,\, .
\label{21}
\eeq

In previous work$^{\mbox{\scriptsize\cite{CM90,MSC91}}}$ we studied momentum
distributions in even--even
deformed nuclei. For spin--saturated even--even nuclei the vector momentum
distribution is zero in the intrinsic ground state and only the scalar
momentum distribution remains. To connect with the notation in
Refs.~\cite{CM90,MSC91}
we denote by $n(\bd{p})$ the contribution from the even--even core to
$\overline{M}(\bd{p})$, and by $n_{\lambda}(p)$ its multipoles. The
contribution from
the odd--nucleon is denoted by $n^{\mbox{\scriptsize odd}}(\bd{p})$

\beq
\overline{M}(\bd{p})=n(\bd{p})+
	n^{\mbox{\scriptsize odd}}(\bd{p}) =
		\sum_{\lambda=\mbox{\scriptsize even}}
		P_{\lambda}(\cos \theta_p)\left[
		n_{\lambda}(p)+
		n_\lambda^{\mbox{\scriptsize odd}}(p)\right]
	\equiv \sum_{\lambda=\mbox{\scriptsize even}}P_\lambda (\cos \theta_p)
		\overline{M}_\lambda (p) \, .
	\label{22}
\eeq

A deformation parameter in momentum space $\beta^p$ was introduced in
Refs.~\cite{CM90,MSC91} defined in analogy to the
standard$^{\mbox{\scriptsize\cite{Ra87}}}$ definition of the
mass quadrupole deformation parameter $\beta^r$,

%\beq
%\beta^r=\sqrt{\frac{4\pi}{5}}\frac
%		{\int \rho(\bd{r})r^2P_2(\cos\theta_r)d\bd{r}}
%		{\int \rho(\bd{r})r^2 d\bd{r}}
%	\label{23}
%\eeq

\beq
\beta^p=\sqrt{\frac{4\pi}{5}}\frac
		{\int n(\bd{p})p^2P_2(\cos\theta_p)d\bd{p}}
		{\int n(\bd{p})p^2 d\bd{p}}
	\label{24}
\eeq
For odd--A nuclei the same definition applies replacing $n(\bd{p})$ by
$\overline{M}(\bd{p})$.

In what follows we show scalar and vector intrinsic momentum
distributions of $^{21}$Ne and $^{37}$Ar. The results presented have been
obtained using the SKA effective interaction$^{\mbox{\scriptsize\cite{Ko76}}}$
and the
McMaster version
of the deformed Hartree--Fock code$^{\mbox{\scriptsize\cite{VL78}}}$ that
follows closely the
procedure of Ref.~\cite{VB72}. The HF results for binding energies as well as
for
the charge and mass quadrupole moments and r.m.s. radii are summarized in
table~I. Also given in the table are the results for the even--even isotopes
$^{20}$Ne and $^{36}$Ar. Important differences between the structures of Ne
and Ar isotopes are that the former is prolate while the latter is oblate
and that the unpaired neutrons in $^{21}$Ne and in $^{37}$Ar sit in orbitals
with $\Omega_k=3/2$ and with $\Omega_k=1/2$, respectively, while both nuclei
have $J=3/2$ in their ground states. Also given in table~I are the moments
of inertia and the decoupling parameter (defined as in Ref.~\cite{Mo86}).
It can be checked that with these parameters the angular momentum
sequences of the ground state bands in $^{21}$Ne and in $^{37}$Ar are
well described$^{\mbox{\scriptsize\cite{Edu93}}}$ when compared to
experiment$^{\mbox{\scriptsize\cite{LS78}}}$.
The moments of inertia have been
calculated using the cranking formula.

In Figure~1 we show in three--dimensional plots the contributions to the
scalar momentum distribution in $^{21}$Ne (top) and $^{37}$Ar (bottom)
from the core (left) and from the odd nucleon (right).
As already mentioned, the odd nucleons in
$^{21}$Ne and $^{37}$Ar have $\Omega_k=3/2$ and $1/2$, respectively. Clearly
the odd nucleon contributions are quite different in $^{21}$Ne and in
$^{37}$Ar, this is due to the different $K$ values and mean field deformations.
We shall come back to this point later on. It is also interesting to remark
the difference between odd nucleon and core nucleons contributions. An
important
difference between the momentum distribution of the core ($n(\bd{p})$) and the
momentum distribution of the odd nucleon ($n^{\mbox{\scriptsize odd}}(\bd{p})$)
is that in the first case the monopole term, $n_0(p)$, is dominant while in the
second the $\lambda =2$ and higher multipoles are comparable to the monopole
term. In Figure~2 we show the $\lambda=0,2$ multipoles of the even--even core
of Ne and Ar. It is worth pointing out that the isotropy condition found
in previous work on even--even deformed nuclei$^{\mbox{\scriptsize
\cite{CM90,MSC91}}}$ is also met here for
the even--even cores, in the sense that when we compute the deformation
parameter $\beta^p$ we get $\beta^p \approx 0$ (see table~II), whereas the
corresponding parameter in $r$--space is large, consistently with
the experimental deformation.
Experimentally$^{\mbox{\scriptsize\cite{Ra87,Ra89}}}$ the
quadrupole moments are $51.4\pm 3.8$, $58.5\pm 2.6$,
and $-38.5\pm 21$ fm$^2$, which correspond to $\beta^r$ values of
$0.45\pm 0.03$, $0.51\pm 0.02$, and $-0.15\pm 0.08$,
for $^{21}$Ne, $^{20}$Ne, and $^{36}$Ar, respectively.

Actually, as seen in table~II, $\beta^p$ is close
to zero not only for the even--even, but also for the odd--A nuclei.
This important property follows from the major shell admixtures in the
single--particle wave functions that result from the self--consistent
mean field$^{\mbox{\scriptsize\cite{CM90,MSC91}}}$. Interferences between
different major shells
have opposite sign in $r$--space and in $p$--space, and make it possible
to reach configurations with large equilibrium deformations and minimum
kinetic energies. To illustrate the role of N--admixtures in momentum space
we show on the right hand side of Fig.~2 the results obtained with the Nilsson
model when the Nilsson hamiltonian$^{\mbox{\scriptsize\cite{Ni55}}}$
is diagonalized in the entire basis space
(N$\leq 7$) and when it is diagonalized within each major shell subspace
(the $\chi$, $\mu$ parameters have been chosen as in Ref.~\cite{RS80}).
Clearly, the results for $n_2(p)$ are
substantially different in both cases. When $\Delta$N admixtures are taken
into account the results are similar to the ones obtained in DDHF: the
oscillations in $n_2(p)$ are such as to minimize the $\beta^p$ value. When
$\Delta$N admixtures are neglected the $\lambda=2$ multipole of the density
in momentum space has the same shape as the $\lambda=2$ multipole of the
density in $r$--space, and the $\beta^p$ value turns out to be the same
as the deformation parameter in coordinate space $\beta^r$ (see table~II).
A similar situation is found for the $\lambda=2$ multipoles of the odd nucleon
momentum distributions $n^{\mbox{\scriptsize odd}}(\bd{p})$.

The vector momentum distribution components
$\widehat{M}_z(\bd{p})$ and $\widehat{M}_\bot(\bd{p})$
are also shown on 3--dimensional plots in Figure~3, where
the results on the top correspond to $^{21}$Ne and the results
underneath correspond to $^{37}$Ar. As
seen in this figure, the momentum distributions in $^{37}$Ar have a richer
structure than those in $^{21}$Ne. This particularly rich intrinsic structure
is characteristic of deformed odd--A nuclei with $K=\frac{1}{2}$.
A characteristic of nuclei with $K > \frac{1}{2}$ is the nodal line along
the axis $p_\perp =0$ (see the top plots of Fig.~3). As a rule, the structure
tends
to become simpler for the higher $K$ values. As an example, we show
in Figure~4 the momentum distribution of $^{25}$Mg, that has
$K=\frac{5}{2}$. In this case
($\Omega_k=K=j$), $\widehat{M}_z(\bd{p})=n^{\mbox{\scriptsize odd}}(\bd{p})$
and
$\widehat{M}_\bot(\bd{p})=0$.
The function $n^{\mbox{\scriptsize odd}}(\bd{p})$, represented
in the top of Fig.~4, shows a simple structure corresponding to a $99\%$
$d_{5/2}$ wave function. At the bottom of Fig.~4, one can also see the core
momentum distribution in $^{25}$Mg which is very similar to that of $^{21}$Ne.

To understand the structure of these figures it is convenient to express
the densities in momentum space in terms of the spherical components of
the single--particle wave functions. Using Eqs.~(\ref{9})--(\ref{12}) one can
show that

\beqa
\overline{M}_\lambda(p)=n_\lambda(p)+n_\lambda^{\mbox{\scriptsize odd}}(p)
     & &   =\sum_i 2\left[v_i^2+\delta_{i,k}\left(\frac{1}{2}-v_i^2
		\right)\right]
				\sum_{\ell j}\sum_{\ell' j'}
		   \frac{1}{4\pi}
		   \Theta_{i,\lambda}^{\ell j,\ell' j'}(p)
	\nonumber \\
	& & \times
	\seisj{j}{j'}{\lambda}{\ell'}{\ell}{1/2}
(-1)^{\Omega_i+1/2}\langle j\,\, \Omega_i \,\,j'\,\, -\Omega_i|\lambda \,\,
0\rangle
\label{25}
\eeqa

\beq
\widehat{M}_z(p_\perp ,p_z)=\frac{\sqrt{6}}{4\pi}
		\sum_{L=\mbox{\scriptsize even}}
		\sum_{\lambda =\mbox{\scriptsize odd}}
	\hat{\lambda}\hat{L}P_L(\cos\theta_p)
	\tresj{1}{L}{\lambda}{0}{0}{0}
	\Pi^0_{L,\lambda}(p)
	\label{26}
\eeq

\beq
\widehat{M}_\perp (p_\perp ,p_z)=-\frac{2\sqrt{3}}{4\pi}
		\sum_{L=\mbox{\scriptsize even}}
		\sum_{\lambda =\mbox{\scriptsize odd}}
	\frac{\hat{\lambda}\hat{L}}{\sqrt{(L+1)L}}
	P^1_L(\cos\theta_p)
	\tresj{1}{L}{\lambda}{1}{-1}{0}
	\Pi^0_{L,\lambda}(p)
	\label{27}
\eeq
where we have introduced the momentum dependent density

\beq
\Pi^0_{L,\lambda}(p)=\sum_{\ell j}\sum_{\ell' j'}
		\Theta_{k,L}^{\ell j ,\ell' j'}(p)
	\nuevej{j'}{\lambda}{j}{\ell'}{L}{\ell}{1/2}{1}{1/2}
	(-1)^{\ell'+j'+K}
		\langle j\,\, K\,\, j'\,\, -K|\lambda \,\,0\rangle
	\label{28}
\eeq
with

\beq
\Theta_{k,L}^{\ell j ,\ell' j'}(p)=\tilde{\phi}_k^{\ell j}(p)
		\tilde{\phi}_k^{\ell' j'}(p)
		\hat{\ell}\hat{\ell'}\hat{L}\hat{j}\hat{j'}
		\tresj{\ell}{\ell'}{L}{0}{0}{0}
	\label{29}
\eeq
We use the notation $\hat{a}=\sqrt{2a+1}$.
Eq.~(\ref{29}) also applies to $\Theta_{i,\lambda}^{\ell j ,\ell' j'}(p)$ with
$k$ replaced
by $i$ and $L$ by $\lambda$.
$\tilde{\phi}_k^{\ell j}(p)$ ($\tilde{\phi}_i^{\ell j}(p)$) are
the spherical components of the odd
nucleon (core nucleons) wave function defined in Eqs.~(\ref{12})
and (\ref{13}). Note that
under interchange of $p_z$ by $-p_{z}$, i.e., reflection through the plane
perpendicular to the symmetry axis, $\overline{M}(\bd{p})$ and
$\widehat{M}_z(\bd{p})$ are unchanged, whereas $\widehat{M}_\perp (\bd{p})$
changes sign.

The fact that
each individual single--particle wave function contains in general more than
one
$\ell j$ component, makes it possible to have contributions to
Eqs.~(\ref{25})--(\ref{28})
coming from several $\ell j$--components, as well as contributions from
interferences between different $\ell j$ waves. This is an important feature
that would not be possible in the spherical limit, i.e., within the extreme
spherical shell model. The lower the $K$ value is, the lower $\ell j$--values
are allowed in the single--particle wave functions, producing a richer
structure
in the momentum distributions (compare the results for $^{25}$Mg,
$^{37}$Ar and $^{21}$Ne in
Figs.~1--4). In these sd--shell nuclei the dominant $\ell j$ components
of the outermost nucleons are $\ell =0$ and $\ell =2$, though higher components
are also present in our single--particle HF wave functions. In particular,
the $\ell =0$ wave is allowed for the $K=1/2$ wave function while
it is not for $K=3/2, 5/2$, giving rise to  the richer structure observed in
$^{37}$Ar.

%%%%%%%%%%%%%%%%%%%%%%%%%%%%%%%%%%%%%%%%

\section*{III. Momentum Distribution Spin Matrix in Laboratory Frame.
Polarization Considerations}

To study spin dependent momentum distributions in laboratory frame we need
to consider polarized nuclei. We consider a nucleus in its ground state with
angular momentum $J$ completely polarized in a given direction $\bd{P}^*$
defined by the angles $\Omega^*=(\theta^*,\phi^*)$ in the laboratory frame.
The two by two matrix

\beq
M_{\sigma \sigma'}^{(J)}(\bd{p},\Omega^*)
		=\langle J J(\Omega^*)|a^+_{\bd{p}\sigma}
			a_{\bd{p}\sigma'}|J J(\Omega^*)\rangle
	\label{30}
\eeq
measures the probability to find a nucleon with momentum $\bd{p}$ and spin
projections $\sigma$, $\sigma'$ in the laboratory frame. Our task here is
to relate momentum distributions that may be observed in the laboratory to
the intrinsic momentum distributions studied in the previous section. For
this purpose it is also useful to consider partial contributions to
Eq.~(\ref{30})
in addition to the total momentum distribution above defined. We note
that Eq.~(\ref{30}) can also be written as

\beq
M_{\sigma \sigma'}^{(J)}(\bd{p},\Omega^*)
	=\sum_{R}\langle J J(\Omega^*)|a^+_{\bd{p}\sigma}|
			R\rangle \langle R|
			a_{\bd{p}\sigma'}|J J(\Omega^*)\rangle
	\label{31}
\eeq
where the sum is carried out over a complete set of states $R$ of the residual
system with A-1 nucleons. In coincidence reactions of the type $(\hbox{x},
\hbox{x}'\hbox{N})$, where x represents a leptonic or a hadronic projectile
and N a knock--out nucleon detected in coincidence with the scattered
particle $\hbox{x}'$, one can select transitions to discrete states in the
residual
system and, therefore, one can measure partial momentum distributions of the
form

\beq
M_{\sigma \sigma'}^{(J \rightarrow J_R)}(\bd{p},\Omega^*)=
	\sum_{M_R}\langle J J(\Omega^*)|a^+_{\bd{p}\sigma}|
			J_R M_R\rangle \langle J_R M_R|
			a_{\bd{p}\sigma'}|J J(\Omega^*)\rangle \, .
	\label{32}
\eeq

In what follows we study the connection between the momentum distributions
defined by Eqs.~(\ref{30}) and (\ref{32}) and the intrinsic momentum
distributions.

\subsection*{III.1 Total Momentum Distributions}

To compute the total spin dependent momentum distribution defined in
Eq.~(\ref{30})
we first note that the magnetic substate $|J J(\Omega^*)\rangle$ must be
expressed in terms of states referred to the same quantization axis as the
spin components $\sigma$, $\sigma'$. We take this quantization axis
to be the $z$ axis of the laboratory fixed frame,

\beq
|J J(\Omega^*)\rangle= \sum_M {\cal D}^J_{M J}(\Omega^*)|J M\rangle
	\label{33}
\eeq
The convention for rotation operators
and ${\cal D}$ matrices is as in Refs.\cite{Vi66,BS75}.

On the other hand, we have to write the angular momentum eigenstates
$|JM\rangle$
in terms of the intrinsic ground state $\Psi_k$ of the odd--A deformed
nucleus. For that purpose we use the Bohr--Mottelson
factorization approximation$^{\mbox{\scriptsize\cite{BM75}}}$
and we get after some algebra

\beqa
& & M_{\sigma \sigma'}^{(J)}(\bd{p},\Omega^*)=\sum_{\lambda \mu}
	\sqrt{4\pi}\hat{\lambda}\tresj{J}{J}{\lambda}{J}{-J}{0}
	Y_{\mu}^{\lambda}(\Omega^*)\frac{\hat{J}^2}{16\pi^2}
	\int d\omega \sum_{\Sigma \Sigma'}
	{\cal D}^{1/2 \ast}_{\sigma \Sigma}(\omega)
	{\cal D}^{1/2}_{\sigma' \Sigma'}(\omega) \nonumber \\
& & \times
	\left\{\tresj{J}{J}{\lambda}{K}{-K}{0}(-1)^{J-K}
	{\cal D}_{\mu 0}^{\lambda}(\omega)
		\left[\langle\Psi_k|a^+_{\bd{p}\Sigma}
		a_{\bd{p}\Sigma'}|\Psi_k\rangle
	+(-1)^{\lambda}
		\langle\Psi_{\bar{k}}|a^+_{\bd{p}\Sigma}a_{\bd{p}\Sigma'}|
			\Psi_{\bar{k}}\rangle\right] \right. \nonumber \\
& & +
	\left. \tresj{J}{J}{\lambda}{K}{K}{-2K}\left[
	{\cal D}_{\mu 2K}^{\lambda}(\omega)
		\langle\Psi_k|a^+_{\bd{p}\Sigma}
		a_{\bd{p}\Sigma'}|\Psi_{\bar{k}}\rangle
	+(-1)^{\lambda}{\cal D}^{\lambda}_{\mu -2K}(\omega)
		\langle\Psi_{\bar{k}}|a^+_{\bd{p}\Sigma}a_{\bd{p}\Sigma'}|
			\Psi_k\rangle\right]\right\}
		\nonumber \\
& &    \label{34}
\eeqa
The same result is obtained in Projected HF to lowest order in
$\langle J^2\rangle^{-1}$ (see Refs.~\cite{Mo86,Vi66}).
We note that the first two terms within the brackets are
the intrinsic spin dependent momentum distributions $M_{\Sigma \Sigma'}
(\bd{p})$ and $(-1)^{\Sigma-\Sigma'}M^{\ast}_{-\Sigma -\Sigma'}(\bd{p})$,
introduced in Eq.~(\ref{14}) (see also Eqs.~(\ref{3}) and (\ref{4})), whereas
the two last terms are new.
The latter depend only on the intrinsic momentum distributions
$\widehat{M}_z(p_z,p_{\perp})$ and
$\widehat{M}_{\perp}(p_z,p_{\perp})$, which in turn depend only on the odd
nucleon wave
functions (see Eqs.~(\ref{18}) and(\ref{21})). We define
$\overline{M}^{(J)}(\bd{p}, \Omega^*)$ and
$\widehat{M}^{(J)}_{\alpha}(\bd{p}, \Omega^*)$
(with $\alpha=0,\pm 1$) as the
trace and the spherical vector components, respectively, of the
momentum distribution spin matrix in laboratory frame for the fully polarized
nuclear ground
state
\beq
\overline{M}^{(J)}(\bd{p},\Omega^*)=Tr\left(M^{(J)}
	(\bd{p},\Omega^*)\right)
	\label{35}
\eeq
\beq
\widehat{M}_\alpha^{(J)}(\bd{p},\Omega^*)=Tr\left(M^{(J)}
	(\bd{p},\Omega^*)\sigma_\alpha\right) \, .
	\label{36}
\eeq

It is a simple matter to show that the scalar momentum distribution
$\overline{M}^{(J)}(\bd{p},\Omega^*)$ is proportional to the intrinsic
scalar
momentum distribution and that the $\alpha$ components of the vector
momentum distribution depend only on the odd nucleon. Hence,
also in the laboratory frame, the core
nucleons contribute only to the trace $\overline{M}^{(J)}(\bd{p},\Omega^*)$
and not to $\widehat{\bd{M}}^{(J)}(\bd{p},\Omega^*)$.
Taking the trace in Eq.~(\ref{34}) and using Eq.~(\ref{22}) it is easy to show
that

\beq
\overline{M}^{(J)}(\bd{p},\Omega^*)=\sum_{\lambda=\mbox{\scriptsize even}}
	  \overline{M}_\lambda (p)
		P_\lambda(\cos \theta_p^*)
		G(\lambda;JK)
	\label{37}
\eeq
where $\overline{M}_\lambda (p)$ are the multipoles of the intrinsic momentum
distribution
(see Eq.~(\ref{25})), $G(\lambda;JK)$ is a geometrical coefficient

\beq
G(\lambda;JK)=(-1)^{J-K}\hat{J}^2
	\tresj{J}{J}{\lambda}{J}{-J}{0}
	\tresj{J}{J}{\lambda}{K}{-K}{0} ,
	\label{38}
\eeq
and $\theta_p^*$ is the relative angle between the momentum
$\bd{p}$ and the polarization direction $\bd{P}^*$.

The expression for $\widehat{\bd{M}}^{(J)}(\bd{p},\Omega^*)$ is somewhat more
involved, as one gets contributions from all the terms in Eq.~(\ref{34}) that
correspond to different angular momentum projections of the intrinsic odd
nucleon
densities $\widehat{\bd{M}}(\bd{p})$. After a lengthy
but straightforward algebra, using Eqs.~(\ref{3}) and (\ref{4}), we get
\beq
\widehat{M}^{(J)}_\alpha(\bd{p},\Omega^*)=
		\sqrt{6}\sum_{L=\mbox{\scriptsize even}}\sum_{\lambda=
		\mbox{\scriptsize odd}}\sum_\mu
		(-1)^\alpha Y^\lambda_\mu (\Omega^*)Y^L_{\alpha-\mu}(\Omega_p)
		\tresj{1}{L}{\lambda}{\alpha}{\mu-\alpha}{-\mu}
		{\cal L}_{L,\lambda}^{JK}(p)
	\label{39}
\eeq
with
\beq
{\cal L}_{L,\lambda}^{JK}(p)=G(\lambda;JK)
\left\{\Pi^0_{L,\lambda}(p)+(-1)^{J+K}\frac
		{\langle J\,\, K\,\, J\,\, K|\lambda \,\,2K\rangle}
		{\langle J\,\, K\,\, J\,\, -K|\lambda \,\,0\rangle}
		\Pi^{2K}_{L,\lambda}(p) \right\}
	\label{40}
\eeq
and $\alpha=0,\pm 1$. The momentum dependent densities are given by

\beq
\Pi^{2K}_{L,\lambda}(p)=\sum_{\ell j}\sum_{\ell' j'}
	\Theta^{\ell j,\ell' j'}_{k,L}(p)
	\nuevej{j'}{\lambda}{j}{\ell'}{L}{\ell}{1/2}{1}{1/2}
	(-1)^{\ell'}\langle j\,\, K\,\, j'\,\, K|\lambda \,\,2K\rangle
	\label{41}
\eeq
and $\Pi^{0}_{L,\lambda}(p)$ as given in Eq.~(\ref{28}). Note that the
signature
dependent term, $\Pi^{2K}_{L,\lambda}(p)$, only contributes for
$\lambda$--values
satisfying $2K\leq \lambda \leq 2J$, i.e., $\lambda =2J$ when $K=J$.

If we fix the polarization direction as the $z$--axis in laboratory frame,
Eq.~(\ref{39}) reduces to a simpler angular dependence involving the direction
of
$\bd{p}$ only:

\beq
\widehat{M}^{(J)}_{\alpha}(\bd{p},\Omega^*=0)=
		\sqrt{6}\sum_{L=\mbox{\scriptsize even}}
		\sum_{\lambda=\mbox{\scriptsize odd}}
		(-1)^\alpha \frac{\hat{\lambda}}{\sqrt{4\pi}}
		Y^L_{\alpha}(\Omega_p^*)
		\tresj{1}{L}{\lambda}{\alpha}{-\alpha}{0}
		{\cal L}_{L,\lambda}^{J K}(p) \,\, .
	\label{42}
\eeq

We then define $\widehat{M}^{(J)}_l$, $\widehat{M}^{(J)}_s$ and
$\widehat{M}^{(J)}_n$ as
the vector momentum distribution components
along the polarization axis ($l$) and in
the transverse directions ($s$, $n$). Denoting by ($\theta^*_p, \varphi_p^*$)
the direction of $\bd{p}$ in laboratory frame with the $z$--axis parallel to
$\bd{P}^*$ we have

\beq
\widehat{M}^{(J)}_l(\bd{p},\Omega^*=0)=\frac{\sqrt{6}}{4\pi}
		\sum_{L=\mbox{\scriptsize even}}\sum_{\lambda=
		\mbox{\scriptsize odd}}\hat{\lambda}\hat{L}
		P_L(\cos\theta_p^*)
		\tresj{1}{L}{\lambda}{0}{0}{0}{\cal L}_{L,\lambda}^{JK}(p)
	\label{43}
\eeq

\beq
\widehat{M}^{(J)}_s(\bd{p},\Omega^*=0)=\frac{-1}{\sqrt{2}}
	\left(\widehat{M}_1^{(J)}-\widehat{M}_{-1}^{(J)}\right)\equiv
	\cos\varphi_p^*\widehat{M}_t^{(J)}(\bd{p},\Omega^*=0)
	\label{44}
\eeq

\beq
\widehat{M}^{(J)}_n(\bd{p},\Omega^*=0)=\frac{i}{\sqrt{2}}
	\left(\widehat{M}_1^{(J)}+\widehat{M}_{-1}^{(J)}\right)\equiv
	\sin\varphi_p^*\widehat{M}_t^{(J)}(\bd{p},\Omega^*=0)
	\label{45}
\eeq
with the transverse component defined as

\beqa
\widehat{M}^{(J)}_t(\bd{p},\Omega^*=0)=
	       -\frac{2\sqrt{3}}{4\pi}
		\sum_{L=\mbox{\scriptsize even}}
		\sum_{\lambda=\mbox{\scriptsize odd}}
		\frac{\hat{\lambda}\hat{L}}{\sqrt{(L+1)L}}
		P_L^1(\cos\theta_p^*)
		\tresj{1}{L}{\lambda}{1}{-1}{0}
		{\cal L}_{L,\lambda}^{JK}(p) \,\, .
	\label{46}
\eeqa

The momentum distribution in the laboratory frame for the unpolarized nucleus
can be easily obtained by integrating Eqs.~(\ref{37}) and (\ref{39})
over the polarization direction.
Clearly for an {\bf unpolarized} nucleus only the $\lambda=0$ multipole of the
scalar momentum distribution remains:
\beq
\int d\Omega^*\widehat{M}_\alpha^{(J)}(\bd{p},\Omega^*)=0\,\,\, ,
\,\,\,\,\,\,\,\,\,\,\,\,\,\,
\int d\Omega^*\overline{M}^{(J)}(\bd{p},\Omega^*)=4\pi\overline{M}_
						  {\lambda=0}(p)
	\label{47}
\eeq
For $\lambda=0$, Eq.~(\ref{25}) reduces to
\beq
\overline{M}_{\lambda=0}(p)=
n_{\lambda=0}(p)+n_{\lambda=0}^{\mbox{\scriptsize odd}}(p)=
	\frac{1}{4\pi}\sum_i 2 \left[
	v_i^2+\delta_{i,k}\left(\frac{1}{2}-v_i^2\right)\right]
	\sum_{\ell j} |\tilde{\phi}_i^{\ell j}(p)|^2 \,\, .
		\label{48}
\eeq

Hence in the {\bf unpolarized} case one loses not only all of the information
contained in the vector momentum distribution but also important
information contained in the scalar momentum distribution. One misses the
$\lambda \geq 2$ multipoles of the core and odd nucleon momentum densities
that, as discussed in Section II,
carry important information on shell admixtures and on details of
the internal dynamics which are crucial to attain equilibrium in open
shell nuclei.

Now we can easily compare the scalar momentum distribution in laboratory and
intrinsic
frames. Comparison of Eq.~(\ref{22}) to Eq.~(\ref{37}) shows that the
dependence
on the direction of $\bd{p}$
relative to the intrinsic symmetry axis, is replaced
in the laboratory frame by a similar dependence on the direction of $\bd{p}$
relative to the polarization direction, weighted by a geometrical coefficient
$G(\lambda;JK)$. This geometrical coefficient takes the value 1 for $\lambda=0$
and
decreases as $\lambda$ increases ($G(\lambda;JK)=0$ for $\lambda > 2J$).
For $^{37}$Ar and $^{21}$Ne the only non--zero coefficient with
$\lambda >0$ is $G(\lambda=2)=1/5,-1/5$, respectively.

For the vector momentum distribution we also see that the expressions for
longitudinal and transverse components in the laboratory frame involve
the intrinsic multipoles
$\Pi^0_{L,\lambda}(p)$. These expressions are similar to the ones
in the intrinsic frame (compare Eqs.~(\ref{43}) and (\ref{46}) to
Eqs.~(\ref{26}) and (\ref{27})), replacing
the direction of $\bd{p}$ relative to the internal symmetry axis by the
direction of $\bd{p}$ relative to the polarization direction, and weighting
the intrinsic $\lambda$ multipoles with the geometrical factor $G(\lambda;JK)$.
However in the case of the vector components one has additional contributions
for the $\lambda\geq 2K$ multipoles coming from the signature dependent
terms (see Eqs.~(\ref{40}) and (\ref{41})).

%&&&&&&&&&&&&&&&&&&&&&&&&&&&&&&&&&&&&&&&&&&&&&&&&&&&&&&&&&
%&&&&&&&&&&&&&&&&&&&&&&&&&&&&&&&&&&&&&&&&&&&&&&&&&&&&&&&&&

\subsection*{III.2 Partial Momentum Distributions}

%%%%%%%%%%%%%%%%%%%%%%%%%%%%%%%%%%%%%%%%%%%%%%%%%%%%%%%%%%%%%
%%%%%%%%%%%%%%%%%%%%%%%%%%%%%%%%%%%%%%%%%%%%%%%%%%%%%%%%%%%%%%%

In this section we consider partial contributions, as defined in
Eq.~(\ref{32}), to
the spin dependent momentum distribution previously discussed. These partial
contributions play a central role in coincidence $\vec{\hbox{A}}(\vec{
\hbox{x}},\hbox{x}'\hbox{N})\hbox{R}$ experiments where transitions to discrete
states $J_R$ of the residual nucleus are selected. In these processes the
dependence on the nuclear structure of the cross section contains as basic
ingredients the spin dependent transition densities in momentum space
$M_{\sigma \sigma'}^{(J\rightarrow J_R)}$ defined in Eq.~(\ref{32})
(see Refs.~\cite{CDP94,CDP93}).

As seen in previous sections, where we considered total momentum distributions,
only the unpaired nucleon contributes to the
vector momentum distribution. On the contrary,
the scalar momentum distribution
receives contribution from the core nucleons as well as from the unpaired
nucleon, but the latter depends more strongly on
the specific structure of the particular nucleus under consideration. In the
contribution from the core, the $\lambda =0$ multipole (occurring also for
unpolarized nuclei) dominates, while in the contribution from the odd
nucleon the multipoles with $\lambda > 0$ are comparable to the $\lambda =0$
multipole. Thus, it is interesting to focus on transitions in which the
knock--out nucleon is the unpaired one, and the intrinsic structure of the
residual nucleus is basically given by that of the even--even core of the
parent nucleus. This is the case for transitions to the low lying states in the
residual nucleus which are populated when the odd nucleon is knocked out.

We consider a transition from a $100\%$ polarized nucleus in the
direction $\bd{P}^*=(\Omega^*)$ to a discrete state $J_R$ in the residual
nucleus of unobserved polarization. The spin dependent transition density
in momentum space can then be written as

\beqa
& &M_{\sigma \sigma'}^{(J\rightarrow J_R)}(\bd{p},\Omega^*)=
	\sum_{\lambda \mu}\hat{\lambda}^2{\cal D}_{\mu 0}^{\lambda \ast}
		(\Omega^*) \tresj{J}{J}{\lambda}{-J}{J}{0}
				\nonumber \\
& &\times \sum_{M M' M_R} (-1)^{J-M}\tresj{J}{J}{\lambda}{-M}{M'}{\mu}
	\langle J M|a^+_{\bd{p}\sigma}|J_RM_R\rangle
	\langle J M'|a^+_{\bd{p}\sigma'}|J_RM_R\rangle^*\,\, ,
			\label{49}
\eeqa
where we have replaced Eq.~(\ref{33}) into Eq.~(\ref{32}) and used composition
of rotation
matrices. In addition here we specialize to states $J_R$ in the
ground state rotational band ($K^\pi =0^+$) of the residual nucleus. The matrix
elements
entering into Eq.~(\ref{49}) are in this case
\beqa
\langle J M|a^+_{\bd{p}\sigma}|J_RM_R\rangle &=&
	\frac{\hat{J}\hat{J}_R}{8\pi^2\sqrt{2}}
	\sum_{\Sigma}\int d\omega
	{\cal D}_{M_R 0}^{J_R *}(\omega){\cal D}_{\sigma \Sigma}^{1/2 *}(\omega)
	\nonumber \\
&\times & \left\{ {\cal D}_{MK}^{J}(\omega)
		\langle \Phi_k|a^+_{\bd{p}\Sigma}|\Phi_0\rangle +
    (-1)^{J-K}{\cal D}_{M-K}^{J}(\omega)
		\langle \Phi_{\bar{k}}|a^+_{\bd{p}\Sigma}|\Phi_0\rangle \right\}
    \label{50}
\eeqa
with $\Phi_0$ the intrinsic ground state of the even--even core.

Using the $\ell j$--wave expansion of the single--particle wave functions in
the
intrinsic frame (see Eqs.~(\ref{9}) and (\ref{10})), and transforming to the
laboratory frame we
find that
\beq
\langle \Phi_k|a^+_{\bd{p}\Sigma}|\Phi_0\rangle
= \langle \chi_\Sigma |\tilde{\Phi}_k(\bd{p},s)\rangle^* =
\sum_{\ell j}\sum_{m_\ell} \tilde{\phi}_k^{\ell j}(p)
		\langle \ell\,\, K-\Sigma\,\, \frac{1}{2}\,\, \Sigma|j\,\, K\rangle
		Y^{\ell *}_{m_\ell}(\Omega_p) {\cal D}^{\ell *}_{m_\ell\,K-\Sigma}
		(\omega) \, .
	\label{51}
\eeq

A similar expression is obtained for $\langle \Phi_{\bar{k}}|a^+_{\bd{p}
\Sigma}|\Phi_0\rangle $ replacing
$K$ by $-K$ and multiplying by a factor $(-1)^{j-K}$ in Eq.~(\ref{51}).

Upon substitution of Eq.~(\ref{51}) into Eq.~(\ref{50}) and integration over
the direction
of the intrinsic frame we get

\beqa
\langle J M|a^+_{\bd{p}\sigma}|J_R M_R\rangle &=&
		\hat{J}\hat{J}_R \frac{\left(1+(-1)^{J_R}\right)}{\sqrt{2}}
		\sum_{\ell j}\sum_{m_\ell \mu} \tilde{\phi}_k^{\ell j}(p)
		\tresj{J_R}{J}{J}{0}{K}{-K}
\nonumber \\
& \times & Y^{\ell *}_{m_\ell}(\Omega_p)\langle \ell\,\,m_\ell\,\,\frac{1}{2}
\,\,\sigma|
		j\,\,\mu \rangle \tresj{J_R}{J}{j}{-M_R}{M}{-\mu}
		(-1)^{K-M}
	\label{52}
\eeqa
The factor $(1+(-1)^{J_R})$ appears from the sum of the two contributions
in Eq.~(\ref{50}), and
keeps track of the well known property that $0^+$ bands only
contain states with even $J$ values, as is the case for the ground state
bands of even--even axially symmetric nuclei with reflexion symmetry through
a plane perpendicular to the symmetry axis.

Defining scalar and vector transition densities in analogy to the
scalar and vector total momentum distributions,
\beq
\overline{M}^{(J\rightarrow J_R)}(\bd{p},\Omega^*)
	= Tr \left(M^{(J\rightarrow J_R)}
	(\bd{p},\Omega^*)\right)
	\label{53}
\eeq
\beq
\widehat{M}_\alpha^{(J\rightarrow J_R)}(\bd{p},\Omega^*)
	= Tr \left(M^{(J\rightarrow J_R)}
	(\bd{p},\Omega^*) \sigma_\alpha \right)\,\, ,
	\label{54}
\eeq
we readily find after substitution of Eq.~(\ref{52}) into Eq.~(\ref{49}) and
computation
of the traces in spin space:
\beqa
\overline{M}^{(J\rightarrow J_R)}(\bd{p},\Omega^*)&=&
	\sum_{\lambda=\mbox{\scriptsize even}}\frac{\hat{\lambda}}{4\pi}
	P_\lambda(\cos\theta_p^*)\tresj{J}{J}{\lambda}{-J}{J}{0}
	\hat{J}^2(-1)^{J-K}
		\nonumber \\
&\times & \sum_{\ell j}\sum_{\ell' j'}\Theta_{k,\lambda}^{\ell j, \ell' j'}(p)
	\seisj{j'}{j}{\lambda}{\ell}{\ell'}{1/2}(-1)^{K+j+j'-1/2}
		\nonumber \\
&\times & \tresj{J_R}{J}{j}{0}{K}{-K}\tresj{J_R}{J}{j'}{0}{K}{-K}
	\seisj{j'}{j}{\lambda}{J}{J}{J_R}\hat{J}_R^2
		\left(1+(-1)^{J_R}\right)
	\nonumber \\
& &  \label{55}
\eeqa
\beqa
\widehat{M}^{(J\rightarrow J_R)}_\alpha (\bd{p},\Omega^*)&=&\sqrt{6}
	\sum_{L=\mbox{\scriptsize even}}\sum_{\lambda=L\pm 1}\sum_\mu
	(-1)^\alpha Y_\mu^\lambda(\Omega^*)Y_{\alpha-\mu}^L (\Omega_p)
		\nonumber \\
&\times&
	\tresj{1}{L}{\lambda}{\alpha}{\mu-\alpha}{-\mu}(-1)^{J-K}
	\tresj{J}{J}{\lambda}{-J}{J}{0}\hat{J}^2\hat{\lambda}
		\nonumber \\
&\times &  \sum_{\ell j}\sum_{\ell' j'}\Theta_{k,L}^{\ell j, \ell' j'}(p)
	\nuevej{L}{\lambda}{1}{\ell'}{j'}{1/2}
	{\ell}{j}{1/2}(-1)^{\ell'+j'+K}(-1)^{j-j'}
		\nonumber \\
&\times &\tresj{J_R}{J}{j}{0}{K}{-K} \tresj{J_R}{J}{j'}{0}{K}{-K}
	\seisj{j'}{j}{\lambda}{J}{J}{J_R}
		\hat{J}_R^2 \left(1+(-1)^{J_R}\right) \, .
	\nonumber \\
& &  \label{56}
\eeqa

Note that these scalar and vector transition densities are made up
of the same building blocks (the odd nucleon densities
$\Theta_{k,L}^{\ell j, \ell' j'}(p)$ defined in Eq.(~\ref{29}))
as the intrinsic and laboratory scalar
and vector momentum distributions, respectively, and have similar
expressions. The only difference is that for each $\lambda$
multipole the sums over $\ell j,\ell' j'$ are now restricted to angular
momentum values satisfying $|J-J_R|\leq j$, $j'\leq J+J_R$;
$|j-J_R|\leq j' \leq j+J_R$, while in Eqs.~(\ref{25})--(\ref{28})
and (\ref{37})--(\ref{41}) the sums over $\ell j,\ell' j'$
do not have these restrictions.

Furthermore, using the relation

\beqa
& &\sum_{J_R}\hat{J}_R^2\hat{\lambda}^2\left(1+(-1)^{J_R}\right)
	\tresj{J_R}{J}{j}{0}{K}{-K}\tresj{J_R}{J}{j'}{0}{K}{-K}
	\seisj{j'}{j}{\lambda}{J}{J}{J_R}
\nonumber \\
&=&(-1)^{j-j'}\left[(-1)^{2J+\lambda} \langle j\,\,K\,\,j'\,\,-K |\lambda\,\,0
\rangle
		\langle J\,\,K\,\,J\,\,-K | \lambda\,\,0 \rangle \right.
	\nonumber \\
&\times &\left. \delta_{\lambda,\mbox{\scriptsize odd}} (-1)^{J+j'+2K}
		\langle j\,\,K\,\,j'\,\,K | \lambda\,\,2K \rangle
		\langle J\,\,K\,\,J\,\, K | \lambda\,\,2K \rangle \right]
	\label{57}
\eeqa
that we have derived from the relations in Appendix II of Ref.~\cite{BS75},
one can show that

\beq
\sum_{J_R}\overline{M}^{(J\rightarrow J_R)}(\bd{p},\Omega^*)=
	\sum_{\lambda =\mbox{\scriptsize even}}
	P_\lambda (\cos\theta_p^*) n_\lambda^
	{\mbox{\scriptsize odd}}(p) G(\lambda;JK)\,\, ,
	\label{58}
\eeq
which is the odd nucleon contribution to the total scalar momentum
distribution $\overline{M}^{(J)}(\bd{p},\Omega^*)$ (see Eq.~(\ref{37})),
and that

\beq
\sum_{J_R}\widehat{M}_\alpha^{(J\rightarrow J_R)}(\bd{p},\Omega^*) =
		\widehat{M}_\alpha^{(J)}(\bd{p},\Omega^*)
	\label{59}
\eeq
We would like to remark that in Eqs.~(\ref{58}) and (\ref{59}) the sum over
$J_R$ runs over states belonging
to the ground state band in the residual nucleus. Therefore, these equations
tell that measuring the
transition densities to each state in the ground state band one can map out
entirely the intrinsic momentum distribution spin matrix of the odd nucleon.
In addition, Eq.~(\ref{59}) tells us that all possible information contained in
the vector momentum distribution of the polarized target nucleus can be
obtained by measuring the transition densities to the ground state band in the
residual nucleus. The same is true for the odd nucleon contribution to the
scalar momentum distribution (see Eq.~(\ref{58})), which as noted
above is more interesting than the core contribution in the sense that it
depends more on the nuclear structure and on the polarization. Obviously,
transitions
to higher excited states will bring information on momentum distributions in
two quasiparticle excitations, vibrational excitations,..., but will not add
additional information on the spin dependent momentum distribution of the
polarized
nuclear target.

As already mentioned, the only difference between the partial and total
momentum
distributions in laboratory frame is the restriction imposed by the $J_R$
value. This
restriction however, may cause the transition densities to have quite different
structures depending on the $J_R$ value and on the particular
nucleus considered. This is shown in Figures~5 and 6 where we represent
$\overline{M}^{(J\rightarrow J_R)}(\bd{p},\Omega^*)$,
$\widehat{M}_l^{(J\rightarrow J_R)}(\bd{p},\Omega^*=0)$, and
$\widehat{M}_t^{(J\rightarrow J_R)}(\bd{p},\Omega^*=0)$
for $J_R=0$ and $J_R=2$, respectively. The longitudinal
($\widehat{M}_l^{(J\rightarrow J_R)}$)
and transverse ($\widehat{M}_t^{(J\rightarrow J_R)}$) components (with respect
to the polarization direction) of the vector transition densities are
defined in analogy to $\widehat{M}_l^{(J)}$ and $\widehat{M}_t^{(J)}$ in
Eqs.~(\ref{43})--(\ref{46}). Fig.~5 is for the transition from the ground
state in $^{21}$$\stackrel{\longrightarrow}{\mbox{Ne}}$ to the
ground state in $^{20}$Ne. The scalar
transition density is shown on top, and the longitudinal and transverse
components of the vector transition density are shown below. For this
transition ($J\rightarrow J_R=0$)
the momentum dependence of the transition densities does not
change appreciably in going from $^{21}$Ne to $^{37}$Ar, only their
strength change. Thus, Fig.~5 also serves to represent the transition
densities for the ground state in
$^{37}$$\stackrel{\longrightarrow}{\mbox{Ar}}$ to the ground state
in $^{36}$Ar. For this purpose one has to multiply the results shown in
Fig.~5 by a factor $\approx 14$, which is the ratio between the $d_{3/2}$
strength in the odd nucleon wave functions of $^{37}$Ar and $^{21}$Ne.

On the contrary, the momentum dependence changes much in going from
$^{21}$$\stackrel{\longrightarrow}{\mbox{Ne}}$
to $^{37}$$\stackrel{\longrightarrow}{\mbox{Ar}}$
when considering the transition to the first $2^+$
states of their respective residual nuclei ($^{20}$Ne and $^{36}$Ar). This
is clearly seen in Fig.~6, where the plots on the left correspond to
$^{21}$$\stackrel{\longrightarrow}{\mbox{Ne}}\,
(3/2^+)\rightarrow ^{20}\hbox{Ne}\, (2_1^+)$ and the plots on
the right correspond to
$^{37}$$\stackrel{\longrightarrow}{\mbox{Ar}}\,
(3/2^+)\rightarrow ^{36}\hbox{Ar}\, (2_1^+)$. Scalar momentum
distributions are shown on the top and longitudinal ($l$) and transverse ($t$)
components of the vector momentum distributions for these transitions are
shown below. As already noted when comparing the intrinsic momentum
distribution for $^{21}$Ne and $^{37}$Ar in section II, in the case of
$^{37}$Ar the odd nucleon wave function contains mainly a mixture of $s$
and $d$ waves, whereas in the case of $^{21}$Ne it contains mainly $d$
wave. For the transitions $J \rightarrow J_R=0$ only the $d_{3/2}$
wave component contributes in either case, thus resulting in similar
momentum dependence of the transition densities. For the transitions
$J\rightarrow J_R=2$ all positive parity waves with $j=\frac{1}{2},
\frac{3}{2}, \frac{5}{2}, \frac{7}{2}$ may contribute, but the
$j=\frac{1}{2}$ is only present in the odd nucleon of
$^{37}$Ar, not in $^{21}$Ne. This, together with the fact
that also the $\frac{3}{2}, \frac{5}{2}, \frac{7}{2}$
components enter with different amplitudes
in each case, explains why the momentum dependences in Fig.~6 are
different for $^{21}$$\stackrel{\longrightarrow}{\mbox{Ne}}$
and $^{37}$$\stackrel{\longrightarrow}{\mbox{Ar}}$.
Finally we would like to remark
that the transition densities shown in Figs.~5 and ~6 are the dominant
contributions to the total scalar and vector momentum distributions in
$^{21}$$\stackrel{\longrightarrow}{\mbox{Ne}}$ and
$^{37}$$\stackrel{\longrightarrow}{\mbox{Ar}}$. Transitions densities to
higher excited states
$4^+$, $6^+$,... in the ground state band of the residual nuclei are much
smaller than the ones seen in Figs.~5 and ~6.

%%%%%%%%%%%%%%%%%%%%%%%%%%%%%%%%%%%%%%%%%%%%%%%%%%%%%%%%%%%%%%%
%%%%%%%%%%%%%%%%%%%%%%%%%%%%%%%%%%%%%%%%%%%%%%%%%%%%%%%%%%%%%%%%
\subsection*{III.3 Comparison with Spherical Case}
%%%%%%%%%%%%%%%%%%%%%%%%%%%%%%%%%%%%%%%%%%%%%%%%%%%%%%%%%%%%%%%%%%%
%%%%%%%%%%%%%%%%%%%%%%%%%%%%%%%%%%%%%%%%%%%%%%%%%%%%%%%%%%%%%%%%%%%%

It is interesting to compare the results in previous sections
with the ones corresponding to spherical nuclei. The case of spherical
nuclei can be thought of as a limiting case when the deformation
of the mean field goes to zero. In this limit the single
particle wave functions are eigenstates of angular momentum
and the system does not have a preferred direction unless the
nucleus is polarized by an external field. Hence the main two
differences with the deformed case
are that now, i) each single particle wave function has a single
$\ell j$--wave component ($\tilde{\phi}_{\ell j}^i(p)=\delta_{\ell,\ell_i}
\delta_{j,j_i} \tilde{R}_{n_i \ell_i j_i}(p)$), and ii) there is
no distinction between intrinsic and laboratory frames. We denote by
$\{i\} \equiv \{ n_i \ell_i j_i \}$ the set of orbitals filled by
the nucleons of the even--even core and by
$\{ k\} \equiv \{ n_k \ell_k j_k \}$ the odd nucleon
orbital, which has $j_k=J$ and $\ell_k$ fixed by the parity and $J$ values.

The scalar momentum distribution for the fully polarized nucleus is then
given by
\begin{equation}
\overline{M}^{(J) \mbox{\scriptsize sph.}}(\bd{p},\Omega^\star)=
\sum_{\lambda=\mbox{\scriptsize even}} P_\lambda (\cos{\theta^\ast_p})
(\delta_{\lambda,0}\, n^{\mbox{\scriptsize sph.}}(p)+
n_\lambda^{\mbox{\scriptsize odd, sph.}}(p))
	\label{60}
\end{equation}
where ``sph.'' stands for spherical limit;
the odd nucleon multipoles are given by
\begin{equation}
n_\lambda^{\mbox{\scriptsize odd,sph.}}(p)=
\frac{\hat{\ell}_k^2 \hat{\lambda}^2 \hat{j}_k^2}{4 \pi}
|\tilde{R}_{n_k \ell_k j_k}(p)|^2
\tresj{\ell_k}{\ell_k}{\lambda}{0}{0}{0}
(-1)^{j_k+\frac{1}{2}}
\tresj{j_k}{j_k}{\lambda}{j_k}{-j_k}{0}
\seisj{j_k}{j_k}{\lambda}{\ell_k}{\ell_k}{\frac{1}{2}}
\delta_{j_k,J}
\label{61}
\end{equation}
and $n^{\mbox{\scriptsize sph.}}(p)$ is the spherically symmetric momentum
distribution of the core
\begin{equation}
n^{\mbox{\scriptsize sph.}}(p)=n_0(p)=
\frac{1}{4 \pi} \sum_i v_i^2 (2 j_i +1)|\tilde{R}_{n_i \ell_i j_i}(p)|^2
\,\, .
\label{62}
\end{equation}

A thorough study of the properties of $n(p)$ for even--even spherical
nuclei can be found in Refs.~\cite{CMM86,CMMT87}.
The differences between scalar momentum distributions in the spherical and
deformed
case are now made explicit by comparing Eqs.~(\ref{60})--(\ref{62}) with
Eq.~(\ref{37}) (see also Eq.~(\ref{25})). The core has now only a monopole
contribution, while in the deformed case the core has also small multipoles
$n_\lambda(p)$ with $\lambda > 0$, that approximately average to zero
after integration on $p$ (see Fig.~2 and table~II).

For the odd nucleon contribution the sum over $\ell j$, $\ell' j'$
in Eq.~(\ref{28}) is now reduced to a single term
$\ell j =\ell' j'=\ell_k j_k$, with
$\ell_k j_k$ the values corresponding to the spin and parity of the nucleus,
and with $\Omega_k=K=j_k=J$. Therefore, in the
spherical case all the $\lambda$--multipoles have the same $p$ dependence,
whereas
in the deformed case for each $\lambda$ value we may have a different
dependence on $p$. The geometrical coefficient $G(\lambda; J K)$,
which appears in the deformed case from
the transformation between intrinsic and laboratory frames,
is not present in the spherical case. This
reflects the fact that now, ``a priori", there is no internal preferred
direction.

The longitudinal ($l$) and transverse ($t$) components of the vector
momentum distributions are
\begin{eqnarray}
\widehat{M}^{(J) \mbox{\scriptsize sph.}}_{
       \mbox{\scriptsize   $\left\{ \begin{array}{c}
	     l \\ t
	   \end{array} \right\}$    }
}(\bd{p},\Omega^\ast=0)&=&
   |\tilde{R}_{n_k \ell_k j_k}(p)|^2
   \frac{\hat{\ell}_k^2 \hat{j}_k^2 \sqrt{6}}{4 \pi}
\delta_{j_k,J}
\sum_{L=\mbox{\scriptsize even}} \sum_{\lambda= L \pm 1}
\hat{\lambda}^2 \hat{L}^2
\tresj{\ell_k}{\ell_k}{L}{0}{0}{0} \nonumber \\
&&  \hspace{-2.5cm} \times
(-1)^{\ell_k+1}
\tresj{j_k}{j_k}{\lambda}{j_k}{-j_k}{0}
\nuevej{L}{\lambda}{1}{\ell_k}{j_k}{\frac{1}{2}}{\ell_k}{j_k}{\frac{1}{2}}
	   \left\{ \begin{array}{l}
	     P_L(\cos{\theta^\ast_p}) \tresj{1}{L}{\lambda}{0}{0}{0} \\
	     \frac{-P_L^1(\cos{\theta_p^\ast})}{\sqrt{\frac{L(L+1)}{2}}}
	     \tresj{1}{L}{\lambda}{1}{-1}{0}
		     \end{array} \right\}   \label{63}
\end{eqnarray}
where for simplicity we have chosen the polarization direction along
the laboratory $z$ axis. Again, by comparing Eq.~(\ref{63}) to Eqs.~(\ref{43})
and (\ref{46}) (see also Eqs.~(\ref{40}) and (\ref{41})) we see that the
differences with the deformed case are the absence of the geometrical
factor $G(\lambda;J K)$ and the restriction of the sum over
$\ell j$, $\ell' j'$ to a single term with $\ell_k j_k$ dictated by the
spin and parity of the nucleus, and with $\Omega_k=j_k=J$. It is important to
remark that this restriction produces equal $p$--dependence of the various
scalar and vector momentum multipoles.

It is also
interesting to remark the similarity between Eq.~(\ref{63}), that gives
the longitudinal and transverse vector momentum distributions with respect
to the polarization direction, and Eqs.~(\ref{26}) and (\ref{27}), that give
the
intrinsic vector momentum distribution of the deformed nucleus along the
symmetry axis and in the perpendicular plane.
If one restricts Eq.~(\ref{28}) to a single term with $\ell j= \ell' j'=
\ell_k j_k$, and takes $\Omega_k=j_k=J$, the above mentioned
expressions are identical to Eq.~(\ref{63}).
This reflects the fact that for spherical nuclei polarizing  the nucleus is
the only way to define an internal preferred direction for the system.

Expressions for transition densities in this case can also be easily derived.
Obviously in this case to map out the vector momentum distribution
and the odd nucleon contribution to the scalar momentum distribution
one needs only to consider the transition in which the residual
nucleus is left in its ground state ($J_R=0$). It is easy to show
that in this case the vector momentum distribution for the transition
$J \rightarrow J_R=0$ is also given by Eq.~(\ref{63}), and the scalar
momentum distribution is given by the odd nucleon contribution in
Eq.~(\ref{60}). For the case of a spherical nucleus with odd nucleon
in a single $d_{\frac{5}{2}}$ orbital these momentum
distributions have the same structure as the one
shown in the top plot of Fig.~4.

%%%%%%%%%%%%%%%%%%%%%%%%%%%%%%%%%%%%%%%%%%%%%%%%%%%%%%%%%%%%%%%%%%%%%

\section*{IV. Summary and Final Remarks}

%%%%%%%%%%%%%%%%%%%%%%%%%%%%%%%%%%%%%%%%%%%%%%%%%%%%%%%%%%%%%%%%%%%%%%%%

We have studied spin dependent momentum distributions in deformed nuclei on
the basis of the selfconsistent mean field approximation.
The spin degree of freedom is relevant for odd--A nuclei
that can be polarized. We consider the case of axially and reflection
symmetric mean fields and discuss momentum distributions in the nuclear ground
state. We first study spin dependent momentum distributions in the
intrinsic frame for a set of deformed nuclei
in the sd--shell ($^{21}$Ne, $^{37}$Ar, $^{25}$Mg), and then relate them
to momentum distributions in the laboratory frame and
to transition densities in momentum space that can
in principle be measured in one nucleon knock--out reactions. We do
this in a systematic way by decomposing the spin dependent two by two matrix
$M_{\sigma \sigma'}(\bd{p})$ into scalar and vector components in
spin space. We show that the scalar and vector momentum distributions that
can be measured in the laboratory are intimately related to the corresponding
scalar and vector intrinsic momentum distributions. Thus, measuring the former
gives information on the latter.

The study carried out for the selected nuclei
($^{21}$Ne, $^{37}$Ar, $^{25}$Mg) shows that the even--even cores, which
contribute
only to the scalar momentum distribution, are dominated by the
$\lambda=0$ multipole but contain also quadrupole (and higher) multipoles
that are small and tend to average to zero upon integration on $p$. Actually,
this isotropy condition ($\langle p_\perp ^2\rangle = \langle p_z^2\rangle$)
that was already discussed for the case of even--even nuclei in
Refs.~\cite{CM90,MSC91}, is
found here to be also approximately satisfied for the odd--A deformed nuclei.
More involved structure is found in the vector momentum distribution and, to
a lesser extent, in the odd nucleon contribution to the scalar momentum
distribution.

The vector momentum distributions are found to have very rich structures,
particularly for $^{37}$Ar which is the nucleus with lowest $K$ value in its
ground
state. The richness in structure of the vector momentum distribution, as well
as
of the odd nucleon contribution to the scalar momentum distribution,
is found to decrease with increasing $K$ values. Information on this
internal structure can be gained by measuring the transition densities in
momentum
space by one nucleon knock--out reactions from a {\bf polarized} target. It is
also
shown that most of the interesting information can be obtained from transitions
to the lowest states (ground state band) in the residual nucleus. An
application to
$\vec{\hbox{A}}(\vec{\hbox{e}},\hbox{e}'\hbox{n})\hbox{B}$ reaction on
$^{21}$Ne has already been made$^{\mbox{\scriptsize\cite{CDP94}}}$
and applications to the same type of reaction on
$^{37}$Ar and $^{37}$K are now under study$^{\mbox{\scriptsize\cite{Edu93}}}$.
The expressions relating intrinsic and laboratory momentum distributions
have been derived here within a similar philosophy to that in
Refs.~\cite{Mo86,Moya80} concerning the relations between intrinsic and
laboratory multipoles measured by (e,e$'$) reactions.

We have also compared the results on deformed nuclei with results for spherical
nuclei. This comparison shows that in the spherical limit the structure
of vector and scalar momentum distributions is much simpler, even
when we consider a polarized odd--A nucleus. Obviously for unpolarized nuclei,
whether spherical or deformed, only the monopole part of the scalar momentum
distribution enters into the picture; neither the vector momentum
distribution nor
the higher multipoles of the scalar momentum distributions are accessible
in unpolarized nuclei.

The results presented have been obtained with the SKA interaction,
similar results are obtained with other Skyrme type forces. It would be
interesting to check whether the results presented here are modified when
finite range effective interactions, like Gogny
force$^{\mbox{\scriptsize\cite{Gog75}}}$, are used. In principle this
would allow to study effects of short--range dynamical correlations not
included in the present calculations. It would
also be interesting to study the results obtained with relativistic
deformed HF calculations. The latter calculations may be superior to the
non--relativistic ones for the study of spin degrees of freedom. In
particular for odd--A nuclei the relativistic formalism allows to take
into account in a simpler way effects due to time reversal non invariant
terms in the mean field that have not been considered in the present work.
In the future we plan to study these points.

\vspace{1.5cm}
\noindent
{\Large \bf \sc Acknowledgements}
\vspace{0.5cm}

One of us (J.A.C.) wishes to thank the members of CTP (MIT) for their
hospitality and, particularly, T. W. Donnelly for stimulating discussion.
This work has been supported in part by DGICYT (Spain) under
contract No. PB92/0021--C02--01.

\newpage
%definitions
\def\PL{ {\it Phys. Lett.} }
\def\NP{ {\it Nucl. Phys.} }
\def\PTP{ {\it Prog. Theor. Phys.} }
\def\PRL{ {\it Phys. Rev. Lett.} }
\def\PR{ {\it Phys. Rev.} }

\newpage

%%%%%%%%%%%%%%%%%%%%%%%%%%%%%%%%%%%%%%%%%%%%%%%%%%%%%%%%%%%%%%%%%%%%%%%%%%
%  TABLE CAPTIONS

%\clearpage\newpage
%\centerline{\bf Table Captions}
%\begin{enumerate}
%\item[Table I:] Results of DDHF calculations for binding energies, proton and
%mass quadrupole moments, r.m.s. radii, moments of inertia and decoupling
%parameters.

%\item[Table II:] Values of quadrupole deformation parameters in p--space and
%in r--space from Nilsson model calculations with ($\Delta N\neq 0$) and
%without ($\Delta N =0$) major shell admixtures compared to results of
%DDHF calculations.

%\end{enumerate}

%%%%%%%%%%%%%%%%%%%%%%%%%%%%%%%%%%%%%%%%%%%%%%%%%%%%%%%%%%%%%%%%%%%%%%%%%%%

%  FIGURE CAPTIONS

\clearpage\newpage
\centerline{\bf Figure Captions}
\begin{enumerate}

\item[Figure 1:] Contributions from the even--even core (left) and from
the odd nucleon (right) to the scalar momentum distribution in $^{21}$Ne
(top) and $^{37}$Ar (bottom).

\item[Figure 2:] Monopole ($n_0$) and quadrupole ($n_2$) contributions
(in fm$^3$) to the even--even core momentum distributions (normalized to 1)
of $^{21}$Ne and $^{37}$Ar from DDHF calculations and from Nilsson model
calculations. The results obtained with the Nilsson model without major
shell admixtures are labelled with primes.

\item[Figure 3:] Components of the vector momentum distribution parallel and
perpendicular to the symmetry axis for $^{21}$Ne (top) and $^{37}$Ar (bottom).

\item[Figure 4:] Core (bottom) and odd nucleon (top)
momentum distributions in $^{25}$Mg. The vector momentum
distribution has a single component parallel to the symmetry axis that
coincide with the odd nucleon contribution to the scalar
momentum distribution
($\widehat{M}_z(\bd{p})=n^{\mbox{\scriptsize odd}}(\bd{p})$).

\item[Figure 5:] Scalar ($\overline{M}^{(J\rightarrow J_R)}(\bd{p},\Omega^*)$)
and vector components
($\widehat{M}_{l,t}^{(J\rightarrow J_R)}(\bd{p},\Omega^*=0)$)
of the transitions densities in
momentum space (in fm$^3$)
for $J\rightarrow J_R=0$ in $^{21}$Ne. The results for
$^{37}$Ar are similar except for a scale factor $\sim 14$.

\item[Figure 6:] Same as Fig.~5 for the transition $J\rightarrow J_R=2$ in
$^{21}$Ne (left) and in $^{37}$Ar (right).

\end{enumerate}

\newpage

\begin{enumerate}
\item[Table I:] Results of DDHF calculations for binding energies, proton and
mass quadrupole moments, r.m.s. radii, moments of inertia and decoupling
parameters.
\end{enumerate}

\vspace{1.5cm}

\begin{tabular}{ccccccc}
\hline\hline
   Nucleus
   & $B$ {\footnotesize (MeV)}
   & $Q^\pi$ {\footnotesize (fm$^2$)}
   & $Q^M$   {\footnotesize (fm$^2$)}
   & $\langle r^2 \rangle^{\frac{1}{2}}$ {\footnotesize (fm)}
   & ${\cal I}$ {\footnotesize (MeV$^{-1}$)}& $a$
                                                   \\  \hline

$^{21}$Ne & --165.1 &  44.0 &  92.1 & 3.00 & 4.19 &  --      \\

$^{20}$Ne & --155.5 &  44.0 &  86.8 & 3.00 & 4.38 &   --     \\

            \vspace{-4mm}\\

$^{37}$Ar & --312.3 &--43.1 &--74.6 & 3.41 & 1.89 &--1.159 \\

$^{36}$Ar & --301.9 &--43.1 &--84.8 & 3.41 & 2.50 &  -- \\
\hline\hline
\end{tabular}

\newpage

\begin{enumerate}
\item[Table II:] Values of quadrupole deformation parameters in p--space and
in r--space from Nilsson model calculations with ($\Delta N\neq 0$), and
without ($\Delta N =0$) major shell admixtures, compared to results of
DDHF calculations.
\end{enumerate}

\vspace{1.5cm}

\begin{tabular}{|l|cc|cc|cc|cc|}
\hline\hline

	   &\multicolumn{2}{|c|} {$^{21}$Ne} &
	    \multicolumn{2}{|c|} {$^{20}$Ne} &
	    \multicolumn{2}{|c|} {$^{37}$Ar} &
	    \multicolumn{2}{|c|} {$^{36}$Ar} \\ \hline

       &   $\beta^p$ & $\beta^r$ & $\beta^p$ & $\beta^r$
	& $\beta^p$  & $\beta^r$  & $\beta^p$  & $\beta^r$  \\ \hline

DDHF  & 0.041  &  0.402  &  0.050  &  0.405  &  0.014  &  --0.145  &  --0.011
&  --0.170 \\

Nilsson$^{\Delta N \neq 0}$ & 0.022 & 0.445  & --0.002 & 0.487 & 0.028 &--0.138
& 0.012  & --0.171 \\

Nilsson$^{\Delta N=0}$ & 0.232 & 0.232 & 0.234 & 0.234 & --0.062 & --0.062  &
--0.084 & --0.084
  \\ \hline\hline
\end{tabular}

\end{document}